# Sum Capacity Bounds and Design of Suboptimal Codes for Asynchronous CDMA Systems

S. Dashmiz, S. M. Mansouri, A. Najafi, F. Marvasti *IEEE* Senior Member

Advanced Communications Research Institute (ACRI), Department of Electrical Engineering,
Sharif University of Technology, Tehran, Iran.
Email: marvasti@sharif.edu.

*Abstract*—In this paper, we study two issues in asynchronous communication systems. The first issue is the derivation of sum capacity bounds for finite dimensional asynchronous systems. In addition, asymptotic results for the sum capacity bounds are obtained. The second issue is the design of practical suboptimal codes for binary chip asynchronous CDMA systems that become optimal for high Signal-to-Noise (SNR) ratios. The performance of such suboptimal codes is also compared to Gold and Optical Orthogonal codes. The conclusion is that the proposed suboptimal codes perform favorably compared to other known codes for high SNR asynchronous systems and perform more or less the same as the other codes for the low SNR values.

I. INTRODUCTION

The fundamental limits of asymptotic synchronous CDMA systems have been studied in [1-5] by modeling the spreading sequences with random sequences. In our previous papers, we studied bounds on the sum capacity for finite synchronous CDMA systems together with the design of suboptimal codes that approach the sum capacity for highly overloaded CDMA systems at high Signal-to-Noise ratios (SNR) [6-8]. However, the assumption that users are synchronous is not valid for optical networks and the uplink wireless CDMA systems. In this paper, we plan to extend our results for the sum capacity bounds to finite asynchronous CDMA systems. In [9], asymptotic spectral efficiency for a CDMA system with random spreading and linear detector is derived. The derivation involves the chip waveform and the user delay distributions. An interesting observation of [9] is that the spectral efficiency of a chip asynchronous CDMA system is greater than that of the spectral efficiency of a chip synchronous system. Nevertheless, all the studies in this paper are based on the Gaussian distribution assumption for the user input symbols. In [10], the same authors are exploiting the results of [9] to design low complexity statistically efficient linear receivers. In [11], the authors have developed a general asynchronous Welch Bound Equality (WBE) signature sequences that maximize the sum capacity of a CDMA system for real valued input data and signatures.

In this paper, we intend to derive lower and upper bounds for the sum channel capacity of chip asynchronous CDMA systems in finite dimensional case and also we investigate some of the bounds in asymptotic case. Our bounds show that, despite the interference among the users, it is possible to have an asynchronous CDMA system with a large number of users with almost no loss of information at the receiver side. Our main assumption is that it is possible to estimate the time at which users begin to send their data. This is not a very restrictive condition because some known methods such as tracking and acquisition can be applied for estimating the delay of the users.

In our present paper, we also intend to propose a new class of suboptimum codes for asynchronous systems that become optimal at high SNR values. Surprisingly, even when the system has a large amount of delay, suboptimal codes exist for an overloaded CDMA system. The proposed suboptimal codes are compared to the well-known codes for asynchronous CDMA such as Gold and Orthogonal Optical Codes (OOC) [12-14]. Simulation results show that at high SNR values, the proposed suboptimal codes perform much better than the above codes for asynchronous systems but perform more or less the same as the aforementioned codes for the low SNR values.

The organization of this paper is as follows: In the next section, the preliminaries on mathematical model for asynchronous CDMA systems are discussed. A general theorem for computing lower and upper bounds for channel capacity in the noiseless and noisy cases are derived in Sections III and IV, respectively. Furthermore, the asymptotic behaviors of the sum capacity bounds are discussed in the same sections. Suboptimal asynchronous codes are proposed in Section V. Simulation results on the bounds are presented in Section VI. Decoding procedure and the code performances are explained in section VII, and finally, Section VIII concludes the paper.

## II. Preliminaries

In a CDMA system, each user multiplies its data by a signature vector (of fixed or variable length) and sends it through the channel. In the channel, the transmitted signals are compiled and, thus, a sequence of numbers embedded in noise arrives at the receiver side. Different users may begin to send their data at different times due to the nature of multiple access channels, resulting in an asynchronous transmission. We assume that the system is chip-synchronous, i.e., the delay of users is multiple of the chip time. Thus, for a fixed length signature matrix, the channel model is

$$Y[k] = r\frac{1}{\sqrt{m}}\sum_{i=1}^{n} S_i[k-\tau_i]x_i[k-\tau_i] + N[k] \quad (1),$$

where $n$ is the number of users, $m$ is the signature length, $S_i$ is the signature signal of the $i^{th}$ user, which is periodic with period $m$, and $r$ is a gain factor for achieving the desired SNR; additionally, $x_i[k]$ is the input data of the $i^{th}$ user at time $k$, which is constant over a signature duration. In this equation, $\tau_i$ is the delay of the $i^{th}$ user, $N[k]$ is the noise with PDF $f$ at time $k$, $Y$ is the observed vector at the receiver side. Clearly, it can be assumed that $0 \leq \tau_i < m$. For defining the sum capacity, consider a stream of length $Nm$ of the received vector $Y$; denote this section of $Y$ by $Y_N$ and all user input data which contribute to $Y_N$ are represented by $X_N$. If a maximum delay of $\tau_{max}$ is imposed in the system, the sum capacity is defined as

$$C(m,n,\mathcal{I},\mathcal{S},\tau_{max},\eta) = \max_{S_i \in \mathcal{S}^n} \max_{p(x_1)p(x_2)\cdots p(x_n)}$$

$$\lim_{N\to\infty} \frac{\mathbb{E}_{\underline{\tau}}\big(\mathbb{I}(X_N;Y_N)\big)}{N} \quad (2),$$

where the expectation is over all delay vectors $\underline{\tau} = (\tau_1, \tau_2, \ldots, \tau_n)$ with $\max_{i,j}|\tau_i - \tau_j| \leq \tau_{max}$. Here, $\mathcal{I}$ and $\mathcal{S}$ are the sets of the input and signature alphabets, respectively. Also, $\mathbb{I}(X_N;Y_N)$ is the mutual information between $X_N$ and $Y_N$, $p(x_i)$ denotes the probability distribution of the $i^{th}$ user input and $\eta = \frac{m}{n}\text{SNR}$ is the normalized SNR. Equation (2) implies that it is necessary to maximize the average mutual information over all possible delays with respect to all user signature vectors and multiplicative distributions of input data. In the noiseless case, the notation $C(m,n,\mathcal{I},\mathcal{S},\tau_{max})$ is used instead of $C(m,n,\mathcal{I},\mathcal{S},\tau_{max},\eta)$.

In a synchronous CDMA system with $\mathcal{I}$ and $\mathcal{S}$ as the input and the signature alphabet, the channel model is

$$Y = \frac{1}{\sqrt{m}}rAX + N,$$

where $r$ is calculated from the normalized SNR, $\eta$ [8]:

$$r = \left(\frac{mn\eta\sigma_f^2}{tr\,(\mathbb{E}(AXX^*A^*))}\right)^{\frac{1}{2}},$$

Also, define

$$I_{average}(m,n,\mathcal{J},\mathcal{S},\eta) = \mathbb{E}_A\big(I(X;Y|A,\eta)\big),$$

where $I(X;Y|A)$ is the mutual information of a synchronous CDMA with uniform distribution on the input.

In this paper, the goal is to obtain lower and upper bounds for the sum capacity; to this aim, a family of signature vectors called cyclic-vectors is used. A cyclic-vector is a vector of length $m$ with periodic entries. The lower bound of (2) is obtained in two steps. In the first step, instead of maximizing over all the input product distributions, a uniform product distribution on the input vectors is fixed, and in the second step, we take average over a class of special signature vectors instead of finding the signature code that maximizes the sum capacity. We thus get

$$C(m,n,\mathcal{J},\mathcal{S},\tau_{max},\eta) \geq \mathbb{E}_S\left(\lim_{N\to\infty}\frac{\mathbb{E}_\tau\big(\mathbb{I}(X_N;Y_N)\big)}{N}\right),$$

where $X_N$ has the uniform distribution and $S$ represents the set of signature vectors which are also selected randomly from a group of cyclic vectors related to $m$ and $\tau_{max}$ with uniform distribution.

III. General Theorem for Deriving the Asynchronous CDMA Bound From the Synchronous CDMA Bound

The main idea for calculating the lower bound of asynchronous CDMA is to change the problem to a synchronous CDMA system. To this aim, a family of signature vectors called cyclic-vectors is used. By these cyclic codes and using appropriate sections of the bit stream, the problem can easily be transformed to a synchronous CDMA problem.

***Theorem 1***: (General Theorem for transforming the asynchronous case to the synchronous case)

The sum capacity for an asynchronous system is lower bounded by an average mutual information for an equivalent synchronous system; the relation is given by:

$$C(m,n,\mathcal{J},\mathcal{S},\tau_{max},\eta) \geq I_{average}\left(m-\tau_{max},n,\mathcal{J},\mathcal{S},\eta\frac{m-\tau_{max}}{m}\right).$$

The proof is stated in the appendix.

The $I_{average}(m,n,\mathcal{J},\mathcal{S},\eta)$ is calculated in [8] for various input and signature alphabets. Cosequently, the lower bounds can be achieved by the above general theorem. The main point is that, when $\tau_{max} = 0$ which is the synchronous CDMA case; the lower bound is the same as the lower bound which is achieved in [8]. This fact expresses that the above theorem is a generalization of main results in [8].

In the first example, the lower bound is derived for binary input and binary signature in absence of noise.

***Example 1*** *(Binary Noiseless case):*

$$C(m, n, \{\pm 1\}, \{\pm 1\}, \tau_{max}) \geq n - \log \sum_{j=0}^{\lfloor \frac{n}{2} \rfloor} \binom{n}{2j} \left( \frac{\binom{2j}{j}}{2^{2j}} \right)^{m-\tau_{max}}.$$

The proof is stated in the appendix.

If $c(m, n, \mathcal{I}, \mathcal{S}, \tau_{max})$ denotes the capacity per user, i.e., $c(m, n, \mathcal{I}, \mathcal{S}, \tau_{max}) = \frac{1}{n} C(m, n, \mathcal{I}, \mathcal{S}, \tau_{max})$ and if the loading factor $\beta = n/m$ and $\lambda = \tau_{max}/m$ are fixed as $n$ and $m$ and $\tau_{max}$ approach infinity, the following asymptotic result is derived for both input and signature binary in absence of noise:

***Example 2*** (*Asymptotic Binary Noiseless Case*): If $\zeta \geq 0$, $0 \leq \lambda \leq 1$, and $1 - \lambda$ and $\zeta$ are not simultaneously zero, then

$$\lim_{\substack{n/(m \log n) = \zeta \\ (\tau_{max}/m) \to \lambda \\ m \to \infty}} c(m, n, \{\pm 1\}, \{\pm 1\}, \tau_{max}) \geq \min\left\{ 1, \frac{1-\lambda}{2\zeta} \right\}.$$

The proof is provided in the appendix.

The following lower bound is for binary input and quaternary signature alphabet in noiseless channel.

***Example 3*** (*Binary /Quaternary Noisy case*)**:**

$$C(m, n, \{\pm 1\}, \{\pm 1, \pm j\}, \tau_{max}) \geq n - \log \sum_{j=0}^{\lfloor \frac{n}{2} \rfloor} \binom{n}{2j} \left( \frac{\binom{2j}{j}}{2^{2j}} \right)^{2(m-\tau_{max})}.$$

The proof is stated in the appendix.

The following lower bound is for binary input and binary signature alphabet in noisy channel.

***Example 4*** (*Binary Noisy case*)**:** For any positive real number $\gamma$,

$$C(m, n, \{\pm 1\}, \{\pm 1\}, \tau_{max}, \eta)$$

$$\geq -(m - \tau_{max})\left(\gamma \log e - \log(1 + \gamma)\right) - \log\left(\sum_{k=0}^{n} \frac{\binom{n}{k}}{2^n} \left(\sum_{j=0}^{k} \frac{\binom{k}{j}}{2^k} e^{-\frac{2\gamma r^2 (2j-k)^2}{(1+\gamma)m}}\right)^{m-\tau_{max}}\right) \quad (3)$$

where $r = \sqrt{2\eta}$.

The proof is stated in the appendix.

if the loading factor $\beta = n/m$ and $\lambda = \tau_{max}/m$ are fixed as $n$ and $m$ and $\tau_{max}$ approach infinity, the following asymptotic result is derived for both input and signature binary with function $f$ as the pdf of the channel noise:

**Example 5**: For any arbitrary function $q(x)$,

$$\lim_{\substack{n/m=\beta \\ \tau_{max}/m \to \lambda \\ n,m \to \infty}} c(m, n, \{\pm 1\}, \{\pm 1\}, \tau_{max}, f) \geq 1 - \sup_{t \in [0,1]} \left[H(t) + \frac{1-\lambda}{\beta}\left(\mathbb{E}(q(N_1)) + \log \mathbb{E}\left(2^{-q(N_1 - 2\sqrt{t\beta}Z)}\right)\right)\right],$$

where $Z$ is the standard Gaussian random variable.

The proof is stated in the appendix.

Considering $q(x) = -\gamma \log f(x)$, we obtain the following bound for the Gaussian case:

**Example 6** (*Asymptotic Gaussian Noise Lower Bound*):

$$\lim_{\substack{n/m=\beta \\ \tau_{max}/m=\lambda \\ m \to \infty}} c(m, n, \{\pm 1\}, \tau_{max}, f_\sigma) \geq 1 - \inf_{\gamma} \sup_{s \in [0,1]}$$

$$\left[H(s) + \frac{1-\lambda}{2\beta}\left(\gamma \log e - \log\left(1 + \frac{\gamma}{\sigma^2}(\sigma^2 + 4s\beta)\right)\right)\right]$$

The following theorem gives a conjectured upper bound for a noisy channel.

The following theorems are based on the conjecture that the mutual information is maximized for the input vectors with uniform distribution. The conjecture is first introduced in [7] and has been proved in asymptotic case with both input and signature binary in [2].

**Theorem 2** *(General Theorem for a Conjectured Upper Bound of Asynchronous CDMA System):*

$$C(m,n,\mathcal{I},\mathcal{S},\tau_{max}) \leq \mathcal{U}(m,n,\mathcal{I},\mathcal{S}).$$

where $\mathcal{U}(m,n,\mathcal{I},\mathcal{S})$ is the upper bound for synchronous system with given parameters in noiseless case, which is derived in [8].

The proof is stated in the appendix.

The following theorem suggests an upper bound for noisy case in which $\mathcal{S} \subset S^1$, where $S^1$ is the unit circle.

**Theorem 3** *(General Theorem for the Conjectured Upper Bound for an Asynchronous CDMA System in the Noisy Case):*

$$C(m,n,\mathcal{I},\mathcal{S},\tau_{max},\eta) \leq \mathcal{U}(m,n,\mathcal{I},\mathcal{S},\eta).$$

The proof is provided in the appendix.

In the following example the conjecture upper bound is derived for a binary input and signature system in the absence of noise.

**Example 7** (*A Conjectured Upper Bound for the Noiseless Case*):

$$C(m,n,\{\pm1\},\{\pm1\},\tau_{max}) \leq \min\left(n, mh(\tilde{\delta})\right),$$

where $h(f)$ is the differential entropy, and

$$\tilde{\delta}(x) = \sum_{j=0}^{n} \frac{\binom{n}{j}}{2^n} \delta\left(x - \frac{2j-n}{\sqrt{m}}\right).$$

In which $\delta$ is an impulse function.

The proof is provided in the appendix.

The following upper bound is the asymptotic case of example 7.

**Example 8** (*A Conjectured Asymptotic Noiseless Upper Bound*):

If $\zeta \geq 0$, $0 \leq \lambda \leq 1$, $1-\lambda$, and $\zeta$ are not simultaneously zero, then

$$\lim_{\substack{n/(m\log n)=\zeta \\ (\tau_{max}/m)\to\lambda \\ m\to\infty}} c(m,n,\{\pm1\},\{\pm1\},\tau_{max}) \leq \min\left\{1, \frac{1}{2\zeta}\right\}.$$

The proof is provided in the appendix.

*Example 9* (Binary/Ternary with Uniform Distribution Noiseless Case)

If $\mathcal{I} = \{\pm 1\}$ and $\mathcal{S}$ is a set of ternary $AIN^1$

$$C(m,n,\{\pm 1\},\mathcal{S},\tau_{max}) \geq -\log \sum_{k=0}^{n} \binom{n}{k,k,n-2k} \left(\frac{1}{2}\right)^{n+2k} \left(\frac{1}{3^{2k}} \sum_{\alpha+\beta+\gamma=k} \binom{k}{\alpha,\beta,\gamma}^2\right)^{m-\tau_{max}}.$$

The proof is provided in the appendix.

*Example 10* (Binary/Ternary with Uniform Distribution Noiseless Case)

If $\mathcal{I} = \{\pm 1\}$ and $\mathcal{S} = \{1, e^{\pm j2\pi/3}\}$ from Theorem 8 we have

$C(m,n,\{\pm 1\},\mathcal{S},\tau_{max})$

$$\geq -\log \sum_{k_1+k_2+k_3=n} \binom{n}{k_1,k_2,k_3} \left(\frac{1}{2}\right)^{n+k_1+k_2} \left(\frac{1}{3^{k_1+k_2}} \sum_{\alpha+\beta+\gamma=k_1} \binom{k_1}{\alpha,\beta,\gamma}\binom{k_2}{\alpha-\frac{k_1-k_2}{3},\beta-\frac{k_1-k_2}{3},\gamma-\frac{k_1-k_2}{3}}\right)^{m-\tau_{max}}.$$

The proof is provided in the appendix.

*Example 11* (Binary/Real with Uniform Distribution Noisy Case)

If $\mathcal{I} = \{\pm 1\}$ and $\mathcal{S} = \mathbb{R}$ we have

$$C(m,n,\{\pm 1\},\mathbb{R},\tau_{max},\eta) \geq \sup_{\gamma} \left(-(m-\tau_{max})(\gamma \log e - \log(1+\gamma)) - \log\left(\sum_{k=0}^{n} \frac{\binom{n}{k}}{2^n}\left(1+\frac{8k\gamma\eta}{m}\right)^{-\frac{m-\tau_{max}}{2}}\right)\right)$$

The proof is stated in the appendix.

The following lower bound is related to the most general alphabets, in which both input and signature alphabets are real.

---
[1] By AIN we mean algebraic independent numbers over the rational

*Example 12 (Real/Real Noisy Case):*

$$C(m, n, \mathbb{R}, \mathbb{R}, \tau_{max}, \eta) \geq \sup_{\gamma} \left\{ -\left(\frac{m-\tau_{max}}{2}\right)(\gamma \log e - \log(1+\gamma)) - \log\left(\frac{1}{2^{\frac{n}{2}}\Gamma(\frac{n}{2})} \int_0^\infty \left(1 + \frac{4\gamma y \eta}{(1+\gamma)m}\right)^{-\frac{(m-\tau_{max})}{2}} y^{\frac{-n}{2}} e^{\frac{-y}{2}} dy \right) \right\}$$

The proof is stated in appendix.

The lower bound of asymptotic case of example 12, in with both input and signature alphabets are real, is expressed below:

*Example 13 (Real/Real asymptotic Noisy Case):*

$$\lim_{\substack{n/m=\beta \\ \tau_{max}/m=\lambda \\ m \to \infty}} \frac{1}{n} C(m, n, \mathbb{R}, \mathbb{R}, \tau_{max}, \eta) \geq \log e \times \sup_{x \geq 0} \{F(x) - I(x)\}$$

where $F(x) = \frac{-1}{2\beta} \ln(1 + \gamma + 2\gamma \eta \beta x)$ and $I(x) = \sup_t \{xt + \log \sqrt{1-4t}\}$.

For the proof see the appendix.

V. SUBOPTIMAL CODES FOR BINARY ASYNCHRONOUS CDMA

In this section, a class of suboptimal codes for noiseless binary chip-asynchronous CDMA systems is introduced under the assumption that the maximum relative delays among the users are less than a threshold $\tau_{max}$. By suboptimal codes, we mean that in a noiseless environment at the receiver, one can tell exactly which data was sent by each user. Thus, in the noiseless case, if $Y$ is considered as a function of the input data, then this function should be injective (one-to-one) in order to have an error-free communication. For this purpose, we define four kinds of matrices; $A(m, n, s)$, $\widetilde{A}(m, n, s)$, $B(m, n, s)$ and $D(m, n, s)$, where $m$ and $n$ are the sizes of the matrices. Given a column vector of length $m$, by an $s$-circular shift of the vector $u$, we mean that a circular shift of entries of $u$ in upward direction such that the length of the shift is at most $s$. For example, if the entries of $u$ are shifted so that the first entry goes to the $(m-1)^{th}$ entry and the second entry goes to the last entry, a 2-circular shift of $u$ will be achieved. Now, the $s$-rotation of a given matrix $\mathbf{M}$ is a new matrix such that each of its columns is an $s$-circular shift of the corresponding column of $\mathbf{M}$.

Furthermore, let the $\bar{s}$-rotation of $M$ be the matrix obtained exactly like an s-rotation of $M$ with the difference that, after the derivation of the $s$-rotation of $M$, if the first $k$ entries of a column are shifted to the last entries, then their signs are changed. For example, if a 2-circular shift of the first column of $M$ is executed and then the sign of the $(m-1)$th and the last entry of the first column are changed, then a $\bar{2}$-rotation of $M$ will be obtained.

$A(m, n, s)$ is an $m \times n$ binary matrix such that all the $s$-rotations act as an injective function over $\{\pm 1\}^n$. Also, $\tilde{A}(m, n, s)$ is an $m \times n$ $\{0,1\}$ matrix such that all the $s$-rotations act as an injective function over $\{0,1\}^n$. $B(m, n, s)$ is an $m \times n$ binary matrix such that all the $\bar{s}$-rotations act as an injective function over $\{\pm 1\}^n$. $D(m, n, s)$ is an $m \times n$ matrix with $\{0,1\}$ entries such that all the $s$-rotations do not include the zero-vector $\{0\}^m$ and all the one-vector $\{1\}^m$ in its image when its domain is restricted to $\{0,1\}^n \setminus \{0\}^n$ (when the computations are in $\mathbb{GF}_2$).

The next theorem is the main theorem for finding suboptimal codes.

**Theorem 10**: If $A$ is an $A(m - \tau_{max}, n, \tau_{max})$ matrix, then the first $m$ rows of the following infinite dimensional matrix gives a signature matrix which provides $n$ suboptimal codes with $m$ chips when the system has a maximum delay of $\tau_{max}$.

$$\begin{bmatrix} A \\ A \\ \vdots \end{bmatrix}$$

As $\tau_{max}$ decreases, $n$ increases. The next theorem proposes better codes for $\tau_{max} = m$, since for this case the $n$ which is obtained by Theorem 1 will be equal to 0.

**Theorem 11**: If there exists an $A(k, n, k)$ matrix, then for any $i \geq 1$, $n$ suboptimal codes with $m = k \times (n + i)$ chips exist when $\tau_{max} = m$.

In Theorem 1, it is necessary to construct $A$ matrices. For this purpose, a recursive method is introduced based on the $A$, $B$ and $D$ matrices. Theorem 3 gives a method on how to make large $D(m, n, s)$ matrices.

**Theorem 12**: If $W_1$ is a $D(m, n, s)$ matrix, then

$$W = \begin{bmatrix} W_1 & W_1 & 1_{m \times 1} \\ W_1 & 0_{m \times n} & 0_{m \times 1} \end{bmatrix}$$

is a $D(2m, 2n + 1, s)$-matrix. For case $s = m$, $W$ is a $D(2m, 2n + 1, 2m)$-matrix.

For $m = 2^k$, since $\begin{bmatrix}1\\0\end{bmatrix}$ is a $D(2,1,2)$, this theorem reveals that there exists a $D(m, m-1, m)$ matrix. One can easily prove that no $D(m, n, m)$ matrix exists for $n \geq m$.

**Theorem 13**: Assume that $W_1$ and $W_2$ are $A(m, n, s)$, $B(m, k, s)$ matrices, respectively, and $M_1$ and $M_2$ are binary matrices such that $(M_1 + M_2)/2$ is a $D(m, l, s)$ matrix; then,

$$W = \begin{bmatrix} W_1 & W_2 & M_1 \\ W_1 & -W_2 & M_2 \end{bmatrix}$$

is an $A(2m, n + k + l, s)$ matrix. For $s = m$, $W$ is an $A(2m, n + k + l, 2m)$ matrix.

Now, from Theorem 12, it is possible to find large $D$ matrices, especially for the case $m = 2^k$. From this fact together with Theorem 13, one can get large $A$ matrices which results in a large number of suboptimal codes according to Theorem 10. It easy to check that a 2×2 Hadamard matrix is an $A(2, 2, 2)$ matrix.

In what follows, a few examples of suboptimal codes will be given for various values of $\tau_{max}$ for $m = 64$.

**Example 1** (*Suboptimal Codes for $m = 64$ and $\tau_{max} = 16$*):

Computer search confirms that $B(3,2,3)$, $B(6,4,6)$, $B(12,5,12)$ and $B(24,9,16)$ matrices exist. Since there is a $D(3,1,3)$ matrix, Theorem 3 expresses that it is possible to have $D(6,3,6)$, $D(12,7,12)$ and $D(24,15,24)$ matrices. Now, using the above theorems one can recursively construct an $A(48,48,16)$ matrix. Therefore, there are 48 errorless codes for $m = 64$ and $\tau_{max} = 16$.

**Example 2** (*Suboptimal Codes for $m = 64$ and $\tau_{max} = 32$*)**:**

Computer search shows that $B(2,1,2)$, $B(4,2,4)$, $B(8,4,8)$ and $B(16,6,16)$ matrices exist; using the above theorems, one can recursively construct $A(4,4,4)$, $A(8,9,8)$, $A(16,20,16)$ and $A(32,41,32)$ matrices, which means that there are 41 suboptimal codes for $m = 64$ and $\tau_{max} = 32$.

**Example 3** (*Suboptimal Codes for $m = 64$ and $\tau_{max} = 48$*):

The above theorems confirm that there exists an $A(16,20,16)$ matrix, which implies that 20 suboptimal codes are available when $m = 64$ and $\tau_{max} = 48$.

**Example 4** (*Suboptimal Codes for $m = 64$ and $\tau_{max} = 64$*):

Theorem 4 confirms that there is an $A(8,9,8)$ matrix which generates an $A(8,7,8)$ matrix. Theorem 2 implies that 7 suboptimal codes are available when $m = 64$ and $\tau_{max} = 64$.

Also, it is noteworthy that from [15], for $m = 64$ and $\tau_{max} = 0$, there are 193 suboptimal codes. The above results for $m = 64$ are summarized in the following table.

| $\tau_{max}/m$ | 0 | 0.25 | 0.5 | 0.75 | 1 |
|---|---|---|---|---|---|
| $\beta = n/m$ | 3.01 | 0.75 | 0.64 | 0.31 | 0.11 |

Table I. Overloading factors for different normalized maximum relative delays.

**Theorem 14:** Assume that $T_1$ and $T_2$ are $\tilde{A}(m,n,s)$, $A(m,k,s)$ matrices, respectively, and $M_1$ is a $D(m,l,s)$ matrix; then,

$$W = \begin{bmatrix} T_1 & \dfrac{J+T_2}{2} & M_1 \\ T_1 & \dfrac{J-T_2}{2} & 0_{m \times l} \end{bmatrix}$$

is an $\tilde{A}(2m, n+k+l, s)$ matrix. For $s = m$, $W$ is an $\tilde{A}(2m, n+k+l, 2m)$ matrix.

**Example 5** (*Suboptimal Codes for $m = 64$ and $\tau_{max} = 32$ in Optical Case*)**:**

Note that there exist $\tilde{A}(2,2,2)$ and $A(2,2,2)$ matrices. From Examples 2 combined with theorem 14, one can recursively construct a $\tilde{A}(4,5,4), \tilde{A}(8,12,8), \tilde{A}(16,28,16)$ and $\tilde{A}(32,63,32)$. Using a similar theorem to Theorem 10 we can have 63 users in an optical asynchronous system with chip rate 64 and with a maximum delay of 32.

VI. SIMULATION RESULTS

In this section, simulated lower and upper bounds for the sum capacity for noisy and noiseless systems for various values of $\tau_{max}$ and the number of users are presented.

Fig.1 shows that for noiseless CDMA systems with 64 chips, there can be up to about 240 users without any noticeable loss of information when there is no delay. When the maximum delay increases, the lower bound for the capacity decreases. It is intresting that even when $\tau_{max} = 38$, the CDMA system can be overloaded; because for the case of $n = 64$, the lower bound of capacity is very close to 64, i.e., the capacity is close to the maximum, and since the proposed method in deriving the lower bound is based on averaging over all admisible signatures, there exist signature codes near this average with high probability. Also, as before, it can be dedudced that there exist suboptimal codes of lengths $n = 160$, 88, and 20 for $\tau_{max} = 16$, 32 and 48, respectively. In comparison to the values $n = 61$, 41 and 20 found in Examples 1-3, it seems that the proposed construction methods are better when there are large delays in the system. Moreover, it is apparent that the bounds are tight when $\tau_{max} = 0$.

Fig. 2 depicts the asymptotic behavior of the lower bound of the sum capacity versus $\zeta$ when $\lambda$ is equal to 0.5. It can be seen that the lower bounds for the sum capacity approach the asymptotic bound as the signature length goes to infinity.

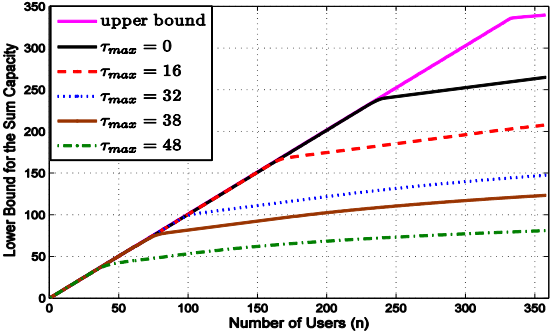

Fig. 1. Lower and upper bounds for the sum capacity versus the number of users for different values of $\tau_{max}$ in the absence of noise when $m = 64$.

The next two numerical evaluations are related to the noisy case. In Fig. 3, it becomes clear again that when the delay increases, the lower bound for the capacity decreases. It is interesting to observe that even for $\tau_{max} = 48$ and $n = 40$, the capacity is more than 0.75 bits per user. One can see that the upper bound is tight w.r.t. the lower bounds especially for the case $\tau_{max} = 0$.

Figure 4 shows that the finite lower bound tends to the asymptotic bound as $m$ approaches infinity.

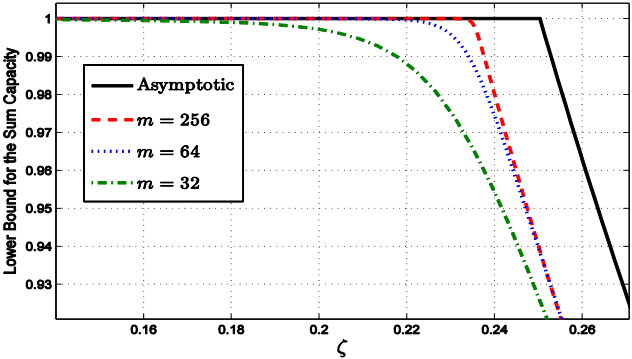

Fig. 2. Lower bound for the sum capacity versus $\zeta$ for the asymptotic as well as finite cases for various values of $m$ and

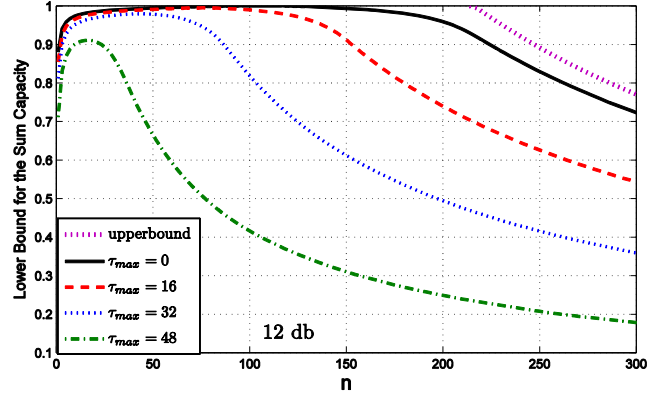

Fig. 3. Lower and upper bounds for the normalized sum capacity versus the number of users for different values of $\tau_{max}$ for the noisy system when $m = 64$ and $E_b/N_0 = 12\ db$.

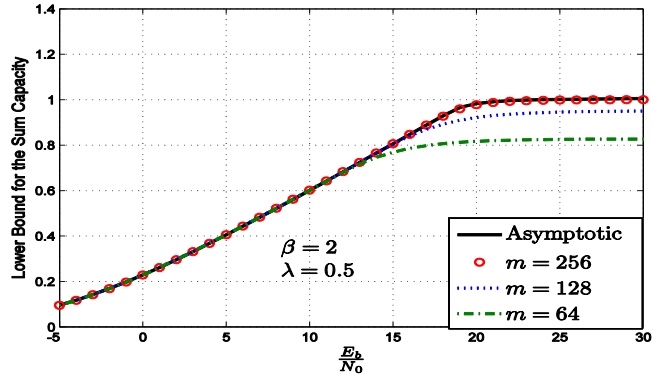

Fig. 4. Asymptotic lower bound for the sum capacity vs. $E_b/N_0$ for several values of $m$ when $\lambda = 0.5$ and $\beta = 2$.

VII. DECODING

As mentioned in Theorem 1, the proposed codes are obtained from the $A$ matrices. We have shown in the proof of Theorem 1 that by choosing a proper window with length $m - \tau_{max}$, the decoding becomes equivalent to that of a synchronous decoder. This decoding scheme is denoted as pseudo ML as long as some information in the time-domain of the signal is ignored, resulting into a suboptimum decoding method. For a more comprehensive comparison, the MAP and a novel method known as iterative soft threshold algorithm are also derived and simulated.

The equation (1) which describes the mathematical model of an asynchronous system can be written in a matrix form as shown in below.

$$y_n = C_p x_n + C_I x_{n-1} + N \qquad (4)$$

It can be seen that in an asynchronous CDMA system, each received vector $y$ is related to both the current and the preceding input vectors denoted by $x_n$ and $x_{n-1}$, respectively. The $C_p$ matrix indicates the signature matrix with each column shifted by its corresponding delay value, and $C_I$ is the interference matrix according to the previous input vector. $N$ is a white Gaussian process with distribution $N(0, \sigma^2)$. The MAP decoding provides optimal result in such systems. This is accomplished by finding the most probable sequence of user inputs for a given received signal:

$$\tilde{x} = \underset{x}{\mathrm{argmax}}\, P(x|y) \qquad (5)$$

Using chain and Bayes rules, we have:

$$P(x|y) = \sum_{x_I} \frac{P(y|x, x_I)}{\sum_x P(y|x, x_I) P(x)} P(x, x_I) \qquad (6)$$

According to (4), $P(y|x, x_I)$ in the above equation can be calculated by the following formula:

$$P(y|x, x_I) = \exp\left(\frac{-1}{2\sigma^2} \|y - C_p x - C_I x_I\|_2^2\right) \qquad (7)$$

The iterative method provides a procedure in order to compensate for the distortion caused by an arbitrary operator $G$, which is fully explained in [16].

$$x_{k+1} = \lambda G x + (I - \lambda G) x_k \qquad (8)$$

$$x_0 = C^T y$$

where $\lambda$ is a constant known as the relaxation parameter and can be adjusted to accelerate the method. It can be shown that under certain conditions, as $k$ goes to infinity, $x_k$ approaches the original signal $x$. The operator $G$ for an overloaded CDMA system is an $N \times N$ matrix defined as follows:

$$G = C^T C, \qquad (9)$$

where $C$ is an $M \times N$ signature matrix. We have observed that a combination of the above mentioned technique with non-linear filters (such as soft thresholding) result in a significant performance for interference cancellation in CDMA overloaded systems. The above-mentioned Soft Thresholding approach applies after each step of the iteration, by setting all the components of the estimated vector that are greater or smaller than two respective threshold values, to specific quantities. Other components would remain unchanged:

$$\psi(x_k) = \begin{cases} 1, & x > \eta \\ -1, & x < -\eta \\ x_k, & O.W. \end{cases} \quad (10)$$

In a more generalized formalism, the mentioned low and high thresholds values can be changed as (5) keeps iterating. The initial threshold values are usually chosen as sufficiently large quantities to avoid the effect of non-linearity in the primary steps of the algorithm, but the values decay exponentially by the number of the current iteration step:

$$\eta_k = \eta_0 e^{-\alpha k}, \quad \alpha > 0 \quad (11)$$

The discussed method is denoted as the Adaptive Soft Thresholding method.

Figure 5 illustrates the performance of the proposed suboptimal codes compared to pseudo Gold sequences. Pseudo Gold codes are a class of binary sequences such that their auto and cross-correlations do not exceed a certain threshold. This threshold has been set to 1 while 7-chip signatures are used for the simulation results. It is shown that the Bit Error Rate (BER) for the proposed codes approaches zero as the SNR increases regardless of the decoding algorithm. Besides, even for low SNR values, the proposed codes have better performance compared to pseudo Gold sequences.

Figure 6 illustrates the performance of the proposed suboptimal codes compared to the Gold sequences. In this case, 9 users are allocated to 32 and 31-chip signatures. Maximum A-Posteriori (MAP) and pseudo ML decoding suffer from computational complexity as long as the size of the look up table required for both algorithms grow exponentially with the number of users. A strategy for handling this problem is to use sphere decoding [17], which reduces the complexity to a manageable level. However, this approach is expected to have a lower performance compared to the pseudo ML or MAP decoding techniques. A modified version of sphere decoding known as Fixed-Complexity Sphere Decoding (FCSD) [17] is employed in this simulation.

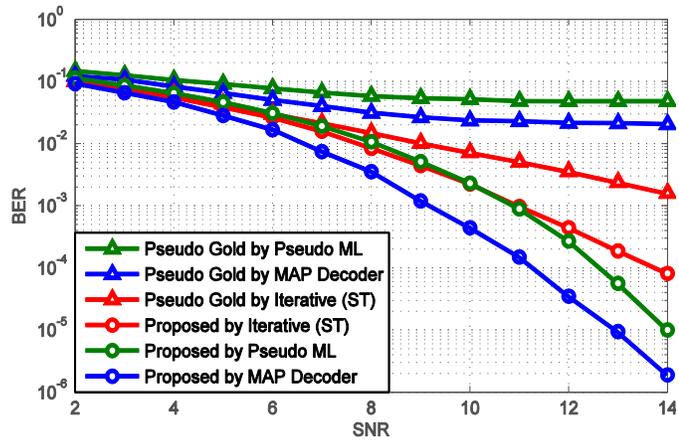

Fig. 5 Performance of the proposed codes vs. pseudo Gold sequences with 4 users, 7-chip signatures and maximum delay of 3 chips.

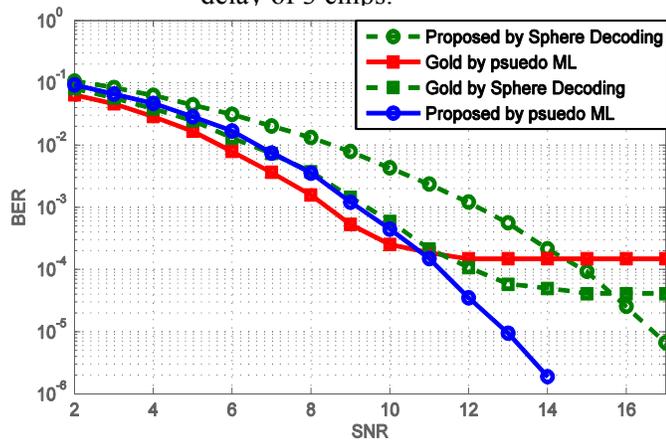

Fig. 6- Performance of the proposed codes vs. Gold sequences with 9 users, 32, 31 chip signatures and maximum delay of 16 chips

From Fig. 6, it can be seen that not only more users are accommodated but also the performance is much better.

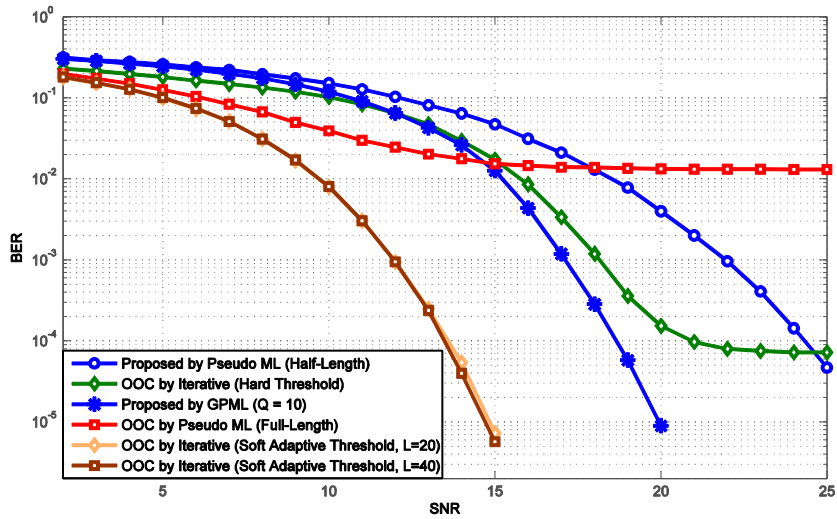

Fig.7 – Comparison 32-chip OOC codes vs. proposed codes for asynchronous optical CDMA with 13 users and maximum delay of 16 chips.

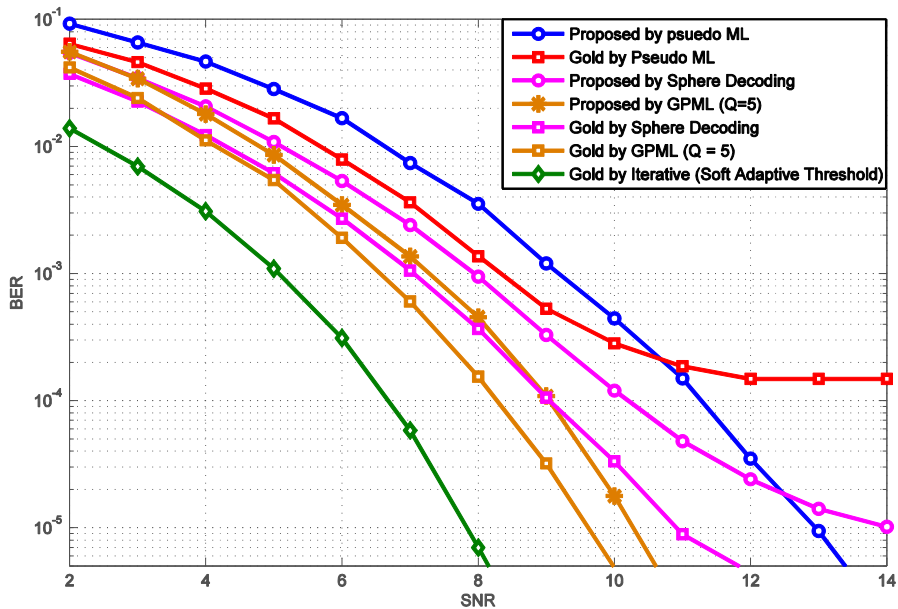

Fig.8 – Comparison 31-chip Gold sequences vs. 32-chip proposed codes for asynchronous CDMA with 9 users and maximum delay of 16 chips.

The previously mentioned pseudo ML method in decoding of asynchronous code division multiplexed data can be generalized into a more reliable method, introduced as Generalized Pseudo ML (GPML). In this case despite the pseudo ML which chose the data with closest last half chips to the received signal $y$, a tunable number of such sequences are kept for further processing. This number is denoted by $Q$ in the simulation results. Let these sequences to be $\{\hat{x}_0, \hat{x}_1, \ldots, \hat{x}_{Q-1}\}$. For each of the mentioned sequences all possible interfering data have been searched to find the most probable pair of main and interfering components for a received vector $y$. According to (4) and (5), the above procedure could be written as follows:

$$\tilde{x}_{GPML} = \underset{\hat{x}_q}{\operatorname{argmin}} \|y - C_p \hat{x}_q - C_I x_I\|_2^2 \qquad \forall x_I, q \in \{0,1,\ldots,Q-1\} \qquad (12)$$

It should be denoted that setting $Q$ to 1 makes the above algorithm equivalent to the aforementioned Pseudo ML method, while setting $Q$ to $2^N$ with $N$ corresponding to the number of users being supported, results in the same performance to that of a MAP receiver. In fact through changing the $Q$ parameter one can manipulate the tradeoff between complexity and performance.

Simulation results demonstrated in Fig.6 and Fig.7 show the performance of proposed codes vs. OOC codes for optical CDMA and Gold sequences for wireless communication respectively. Various method including proposed decoding algorithms have been utilized for decoding the code division multiplexed streams in an AWGN channel. GPML (with $Q = 5$) has shown acceptable performance comparing to iterative adaptive soft threshold method and also the optimum MAP algorithm while reduces the look up table size to $\frac{5}{2^9}$ of its initial value. Iterative method with Soft Adaptive Threshold has outperformed most of other decoding methods specifically when Gold sequences are used as signatures.

VIII. CONCLUSION AND FUTURE WORKS

In this paper, we developed lower and upper bounds for the sum channel capacity of finite asynchronous CDMA systems for both noiseless and noisy cases. The simulations confirm that it is possible to have an overloaded CDMA system without significant loss in the system performance under some fair conditions on the relative delay among the users. Asymptotic lower bounds for the sum capacity are also obtained. In addition, a variety of decoding algorithms have been implemented to demonstrate the performance of the proposed suboptimal codes. It has been shown that proposed codes perform better than other commonly used codes such as Gold and OOC sequences for high SNR values.

For future work, the derivation of the bounds for non-binary real and complex cases, when the delays are not a multiple of chip duration, and near-far effects are good topics to be considered.

## IX. ACKNOWLEDGMENTS

We would like to thank K. Alishahi for his helpful discussions, A. Goli for his help in decoding section, P. Pad for his useful comments, S.Golnarian for some simulations in Section VI and S. M. Sefidgaran for providing Theorem 11.

## X. PROOFS

**Proof of Theorem 1**

As we know, $C(m,n,\mathcal{I},\mathcal{S},\tau_{max},f) \geq \mathbb{E}_S\left(\lim_{N\to\infty} \frac{\mathbb{E}_\tau(\mathbb{I}(X_N;Y_N))}{N}\right)$ and $S$ represents the signature vectors which is also selected randomly from those cyclic vectors with arbitrary element in the first $m - \tau_{max}$ indexed positions and the other entries are a repetition of the first $m - \tau_{max}$ entries. If one considers a window of size $Nm$ of bits, then $Y_N = [\tilde{Y}_1 \cdots \tilde{Y}_N]$ and $X_N = [\tilde{X}_1 \cdots \tilde{X}_N]$; where $\tilde{X}_i$ is all the user's data which contributes to $\tilde{Y}_i$ (each user one data). Now define $\tilde{Y}_{i_1}$ and $\tilde{Y}_{i_2}$ as the same as the definition in Figure 6, we then have

$$\mathbb{I}(X_N;Y_N) = \mathbb{I}(X_N;\tilde{Y}_1 \cdots \tilde{Y}_N) \geq N\,\mathbb{I}(X_1;\tilde{Y}_{1_1})$$

The first inequality is correct because of data processing inequality and the fact that the received vectors $\tilde{Y}_{i_1}$s are independent of each other.

Therefore

$$C(m,n,\tau_{max},f) \geq \mathbb{E}_S\left(\lim_{N\to\infty} \frac{\mathbb{E}_\tau(\mathbb{I}(X_N;Y_N))}{N}\right) \geq \mathbb{E}_S\left(\mathbb{I}(X_1;\tilde{Y}_{1_1})\right) \blacksquare$$

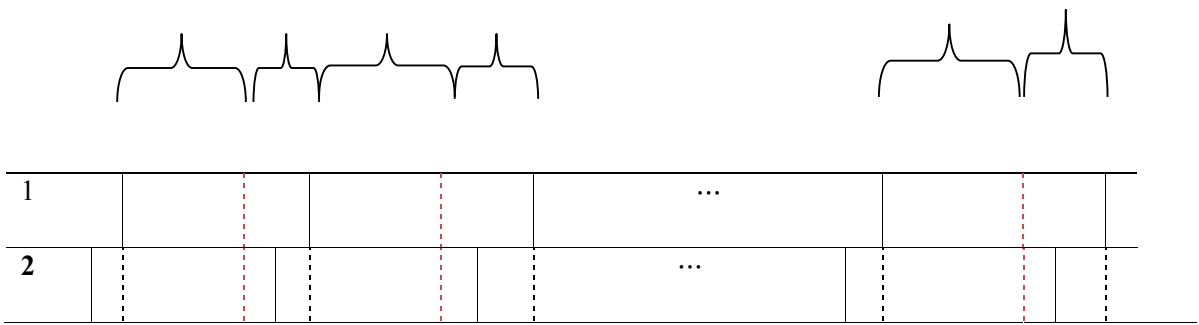

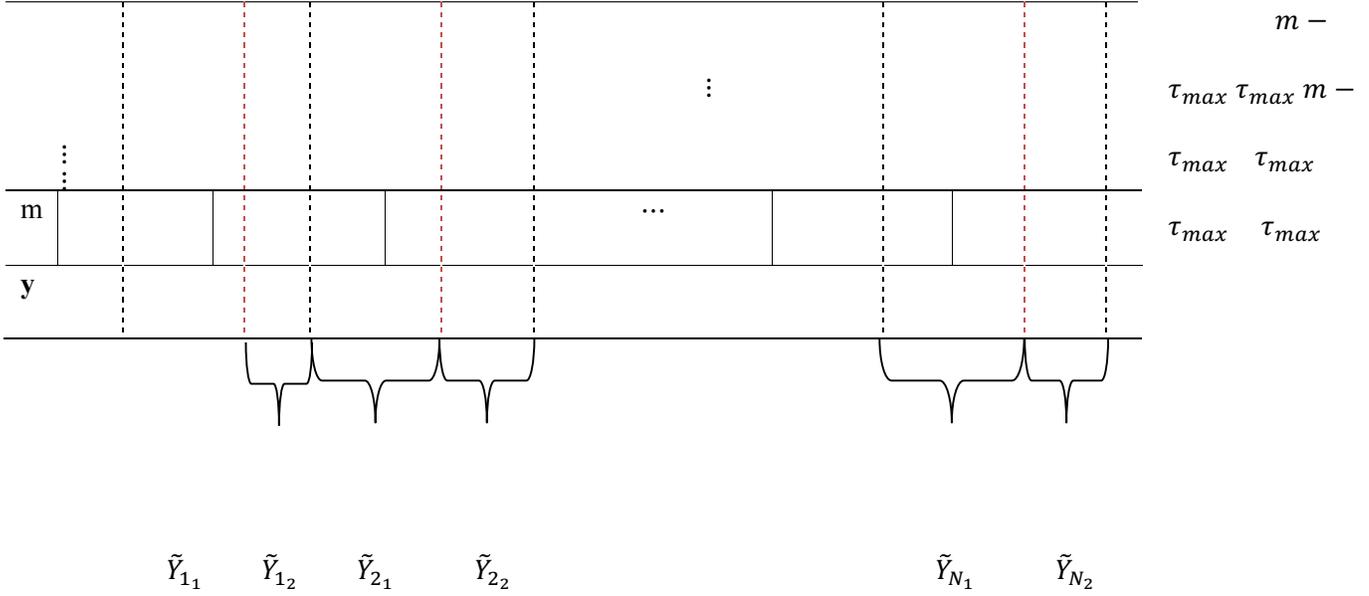

Figure 6. The asynchronous schematic model

**Proof of Theorem 2:** Consider a window (1) with length $m$ as shown in Fig. 6, we have:

$$I(Y, X) = H(Y)$$

Now we have

$$h(Y) = h(Y_1, \ldots, Y_m) \leq \sum_{j=1}^{m} h(Y_j)$$

Based on our conjecture, the mutual information is maximized for the uniform distribution for the input vector. Now note that $Y_j = \frac{1}{\sqrt{m}} \sum_i a_{ji} X_i$. The rest is completely the same as the synchronous case [8].

**Proof of Theorem 3:** It is clear that for binary inputs, we have $C(m, n, \mathcal{I}, \mathcal{S}, \tau_{max}) \leq n\mathbb{H}(\mathcal{I})$. Our conjecture is based on the assumption of $i.i.d.$ user input distribution. Note that $I(X; Y) = \mathbb{H}(Y) - \mathbb{H}(Y|X) = \mathbb{H}(Y_1, Y_2, \ldots Y_m) - m\mathbb{H}(N) \leq m(\mathbb{H}(Y_1) - \mathbb{H}(N))$. When $S = \{s_1, s_2, \ldots, s_q\}$, $\mathbb{H}(Y_1)$ is maximum when we have $u_i$ number of $s_i$ in vector $Y_1$. Let $p$ be the $i.i.d.$ product distribution which maximizes $\mathbb{I}(X; Y)$. Suppose $v_{ij}$ is the number of $s_i \mathcal{I}_i$ in $Y_1 = \frac{1}{\sqrt{m}} \sum_1^n a_{1i} x_i + N_1$. The rest is the same as synchronous case which is derived in [8].

**Proof of Example 7:** From Example 6, we have

$$\lim_{\substack{n/(m \log n) \to \zeta \\ \tau_{max}/m \to \lambda \\ m \to \infty}} c(m, n, \tau_{max}) = \min(1, \frac{mH(\tilde{\delta})}{n = \zeta m \log n}) \leq \min(1, \frac{\log(2\pi e n)}{n = 2\zeta \log n}) \to \min(1, \frac{1}{2\zeta})$$

Therefore, for $\tau_{max} = 0$, combining this with the asymptotic noiseless lower bound, we have the exact capacity for the asymptotic case which is equal to $min\left\{1, \frac{1}{2\zeta}\right\}$.

**Proof of Theorem 10:** Suppose we have a $A(m - \tau_{max}, n, \tau_{max})$ matrix and let $A_1, \cdots, A_n$ be its columns. Now assume $T_1, \cdots, T_n$ are cyclic codes constructed from the vectors $A_1, \cdots, A_n$, respectively. We Put copies of $A_i$ besides each other and consider only the first $n$ entries, i.e., $T_i = [A_i \quad A_i \quad \cdots]$. We wish to prove that $T_1, \cdots,$ and $T_n$ is suboptimal. Note that $\tau_i$ is the delay of the $i$th user. Without loss of generality, suppose that $\tau_n \leq \tau_{n-1} \leq \cdots \leq \tau_1 \leq \tau_{max}$. Consider a window of size $Nm$ bits as the figure shows. Suppose that the window start from the beginning of the first user signature. Now we divide the window into $N$ windows of length $m$ such that the first entry of each window corresponds to the first entry of the first user signature vector. We can divide each $\tilde{Y}_i$ to $[\tilde{Y}_{i_1} \tilde{Y}_{i_2}]$ in such a way that $\tilde{Y}_{i_1}$ has a length of $m - \tau_{max}$. Now consider $\tilde{Y}_{i_1}$ and the corresponding window of all user signatures related to the time of $\tilde{Y}_{i_1}$ for different values of $i$. By the definition of cyclic codes, the parts of user signature vectors in this window are a $\tau_{max}$-rotation of matrix $A$; therefore, by knowing $\tilde{Y}_{i_1}$, we can recognize the signs of signature vectors of users without any error. This implies that the communication is suboptimal.

**Proof of Theorem 11:** Assume that $T = \begin{bmatrix} A \\ \vdots \\ A \end{bmatrix}$, where the number of $A$ matrices are $n + i$. Suppose that $T_1, \cdots, T_n$ are the columns of matrix $T$. We prove that $T_1, \cdots, T_n$ are $n$ suboptimal codes when $\tau_{max} = m$. Consider a window of length $m$ from the stream of input bits. Now, in this window every user contributes by two sections of its signature, corresponding to two different bits of the same user. Since we have only $n$ columns, the position of the beginning of the second bit by each user occurs at most in $n$ intervals $[pk, (p + 1)k]$. Therefore, there exists a number $u$ such that, in the interval $[uk, (u + 1)k]$, no user begins to send his data. However, the window which corresponds to this interval is a $k$-rotation of matrix $A$. Thus, we can recognize the signs of signature users in that interval. Note that the delays are known at the receiver end; thus, by repeating this procedure in the next window of length $m$, we confirm that all the bits can be detected at the receivers..

**Proof of Theorem 12:** Note that, in this case, all the computations are in $\mathbb{GF}_2$. Suppose that the theorem is not true; therefore, there exists an $s$-rotation of $W$ such as $\dot{W}$ and a vector $y$ so that $\dot{W}y$ is zero or 1-vector. Since $\dot{W}$ is an $s$-rotation ($2m$-rotation for $s = m$) of $W$, we can write $\dot{W}$ as the form $\begin{bmatrix} M & T_1 & L_1 \\ M & T_2 & L_2 \end{bmatrix}$ where $M$ and $T_1 + T_2$ are $s$-rotations of $W_1$ and $L_1 + L_2$ is 1-vector. Now write $y$ in the form $\begin{bmatrix} y_1 \\ y_2 \\ 1 \text{ or } 0 \end{bmatrix}$. If the last element of vector $y$ is 0, the terms $My_1 + T_1y_2$ and $My_1 + T_2y_2$ are either zero or 1-vectors. The summation of the two equations confirms that $(T_1 + T_2)y_2$ is either 0 or 1-vector. Since $T_1 + T_2$ is an $s$-rotations of $W_1$, one can conclude that $y_2$ is zero. Using this fact, one can derive that $y_1$ is also zero. Now, if the last element of vector $y$ is 1, the terms $My_1 + T_1y_2 + L_1$ and $My_1 + T_2y_2 + L_2$ are either 0 or 1-vectors. The summation of the two equations confirms that $(T_1 + T_2)y_2$ is either 0 or 1-vector. Since $T_1 + T_2$ is an $s$-rotations of $W_1$, one can conclude that $y_2$ is zero. From this fact, by subtracting the above two equations, the term $L_1 - L_2$ is zero. But we know that $L_1 + L_2$ is 1-vector, which is a contradiction.

**Proof of Theorem 13:** Suppose that the theorem is not true, therefore there exists an $s$-rotation ($2m$-rotation for the case $s = m$) of $W$ such as $\dot{W}$ and a vector $y$ so that $\dot{W}y$ is zero. Since $\dot{W}$ is an $s$-rotation ($2m$-rotation for $s = m$) of $W$, we can write $\dot{W}$ in the form $\begin{bmatrix} M & T & L_1 \\ M & -T & L_2 \end{bmatrix}$, where $M$ and $T$ are $s$ and $\bar{s}$-rotations of $W_1$ and $W_2$, respectively. Now we can write the vector $y$ in the form $\begin{bmatrix} \alpha \\ \beta \\ \gamma \end{bmatrix}$. Now, we have:

$$M\alpha + T\beta + L_1\gamma = 0$$
$$M\alpha - T\beta + L_2\gamma = 0$$

If we sum these equations, we have $M\alpha + \frac{L_1+L_2}{2}\gamma = 0$ ; but $M\alpha$ is a zero or 1-vector when we do the calculations in $\mathbb{GF}_2$. Now, since $\frac{L_1+L_2}{2}$ is an $s$-rotation of $D(m, l, s)$, we have $\gamma = 0$. The above equations show that $M\alpha$ and $T\beta$ are zero. Therefore, $\alpha$ and $\beta$ are also zero because of the definitions for $W_1$ and $W_2$. Hence, $W$ is equal to $A(2m, n + k + l, s)$.

**Proof of Theorem 14:** Suppose that the theorem is not true, therefore there exists an $s$-rotation ($2m$-rotation for the case $s = m$) of $W$ such as $\dot{W}$ and a vector $y$ so that $\dot{W}y$ is zero. Since $\dot{W}$ is an $s$-rotation ($2m$-rotation for $s = m$) of $W$, we can write $\dot{W}$ as the form $\begin{bmatrix} M & T_1 & L_1 \\ M & T_2 & L_2 \end{bmatrix}$ where $M$ and $T$ are $s$ and $\bar{s}$-rotations of $W_1$ and $W_2$, respectively. Now we can write the vector $y$ in the form $\begin{bmatrix} \alpha \\ \beta \\ \gamma \end{bmatrix}$. Then we have:

$$M\alpha + T_1\beta + L_1\gamma = 0$$
$$M\alpha + T_2\beta + L_2\gamma = 0$$

If we sum these equations, we have $\alpha + J\beta + (L_1 + L_2)\gamma = 0$; but $2M\alpha + J\beta$ is a zero or 1-vector when we do the calculations in $\mathbb{GF}_2$. Now, since $(L_1 + L_2)$ is an $s$-rotation of $D(m, l, s)$, we have $\gamma = 0$. The above equation with $\gamma = 0$ and the fact that $(T_1 - T_2)$ is an $A$ matrix indicates that $\beta = 0$. Therefore, $\alpha$ is also zero. Hence, $W$ is an $\tilde{A}(2m, n + k + l, s)$.